\documentclass{aa}
\usepackage{graphicx}

\begin{document}

\title{Periodic variations in the colours of the classical T~Tauri star RW~Aur~A}

\author{P.P.Petrov\inst{1}\fnmsep
	\thanks{on leave from the Crimean Astrophysical Observatory},
	Jaan Pelt\inst{1,2},
	Ilkka Tuominen\inst{1}}

\offprints{P.P.\ Petrov; Peter.Petrov@Oulu.Fi}
\institute {Astronomy Division, P.O.~Box 3000, FIN--90014 University of
Oulu, Finland
\and Tartu Observatory, 61602 T\~oravere, Estonia}

\date{Received ; accepted }

\authorrunning{P.P.\ Petrov et al.}
\titlerunning{Periodic variations in colours of RW~Aur}

\abstract{
The classical T Tauri star \object{RW Aur A} is an irregular variable with a large
amplitude in all photometric bands. In an extended series of photometric
data we found small-amplitude periodic variations in the blue colours of the star,
with a period of $2\fd64$.
The period was relatively stable over several years.
The amplitude of the periodic signal is $0\fm21$ in U--V, $0\fm07$ in B--V,
and about $0\fm02$ in V--R and  V--I.
No periodicity was found in the V magnitude.
The relevance of this photometric period to the recently discovered
periodicity in spectral features of the star is discussed, and the
hypothesis of a hot spot is critically considered.
}
\maketitle

\section{Introduction}


RW Aur is a classical T Tauri star (CTTS) with an unusually strong and variable
emission line spectrum
and large amplitude irregular variations in the brightness as well as in the veiling
of the photospheric spectrum. The star is a resolved triple system
(Ghez et al.~\cite{ghez93}) with RW~Aur~A dominating the brightness in the optical
spectrum. In a recent detailed investigation of RW~Aur~A (Petrov et al.
~\cite{petrov01}, hereafter referred to as Paper I) periodic variations in many
spectral features were found,
with a period within $2\fd6$--$2\fd9$. The period was stable over several years of
observation. A model of non-axisymmetric magnetically channeled accretion
seems to explain most of the observed variations. The asymmetry of the accretion
can be caused either by a close, low-mass, invisible companion
(Gahm et al.~\cite{gahm99}), or by a misalignment of the magnetic
and rotational axes.

The presence of an asymmetric accretion shock at the stellar surface should
modulate the strengths of the \ion{He}{i} and \ion{He}{ii} emission lines
presumably formed in
the post-shock region (Calvet \& Gullbring~\cite{calvet98}), while the hot
photospheric spot (10$^4$ K) below the shock should modulate the brightness
of the star. In case of RW~Aur~A, clear periodic modulations in \ion{He}{i},
\ion{Fe}{ii} and other emissions were observed (Paper~I) which gives a unique
opportunity to study the expected periodic modulation of the brightness.

In this paper we search for periodicities in the variations of the brightness
and the colours of RW~Aur~A, using all the available photometric data
collected over three decades.

\section{Observational data} \label{sec:data}

\begin{figure}
\centerline{\resizebox{8cm}{!}{\includegraphics{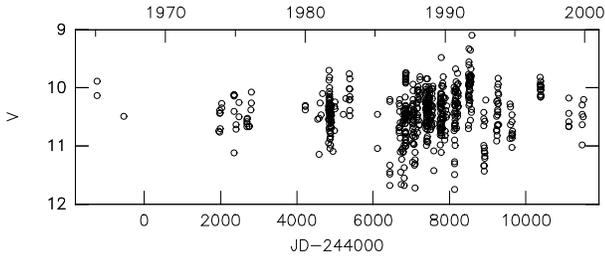}}} 
\caption{Light curve of RW~Aur~A from 1965 to 1999}
\label{mag}
\end{figure}

\begin{figure}
\centerline{\resizebox{6cm}{!}{\includegraphics{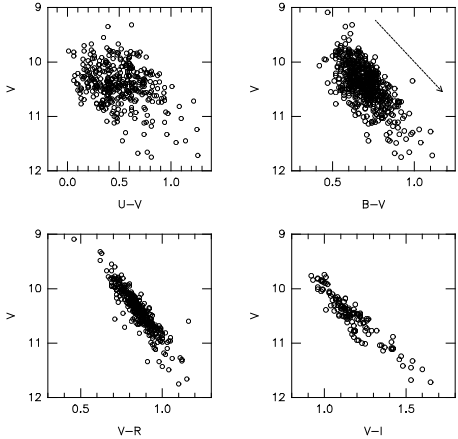}}} 
\caption{Colour--magnitude diagrams. The direction of reddening is shown in
the V {\it versus} B--V diagram according to the mean law of interstellar extinction.
The UBVR bands are in the Johnson system, the I band is in the Cousins
system}
\label{col_mag}
\end{figure}

The data were taken from the catalogue of UBVRI photometry of TTS compiled
by Herbst et al. (\cite{herbst94}). For RW~Aur~A, the catalogue contains 575
observations in V and B--V, and fewer in other colours. The major
part of the data are from the Majdanak Observatory (the ROTOR program by
V.S.Shevchenko's group), and from Van Vleck Observatory (unpublished),
other sources are Kardopolov \& Filip'ev (\cite{kardopol85}),
Rydgren et al. (\cite{rydgren84b}) and Herbst et al. (\cite{herbst83}).
We also used the photometric data of 1996--99, presented in Paper~I.
Since the information about the observational errors is not available
in all cases, we adopt equal weights for all data.
The light curve is shown in Fig.~\ref{mag}, and the colour-magnitude
diagrams are shown in Fig.~\ref{col_mag}

There is a good correlation between V and the colours B--V, V--R and R--I,
with the slopes roughly corresponding to the mean law of interstellar
extinction. It looks very much like the result of variable circumstellar
extinction.
Note, however, that for the spectral type of RW~Aur~A (K1--K4 V)
the normal photospheric colour B--V should be $0\fm86$--$1\fm05$.
With $A_V=0\fm3$ (Paper~I)
the observed photospheric colour B--V is expected to be within $0\fm95$--$1\fm14$.
Most of the observed B--V colours are much bluer.
This effect is also known for other CTTS: the stars typically have closest to
normal photospheric colours when they are at minimum brightness, while becoming
overly blue at maximum light (Vrba et al.~\cite{vrba93}). This is believed
to be due to an additional, variable non-photospheric hot continuum which
also accounts for the veiling of the photospheric line spectrum.
The continuum may belong either to hot photospheric spots at the base of the
accretion columns, or to plage-like regions of hot gas above the
stellar surface. The contribution of the emission lines to the photometric
bands also accounts for a part of the excessively blue colours.

\section{Time series analysis}

\subsection{Feature at $P=2\fd641$}

The compiled data set of RW~Aur~A is very inhomogeneous.
It contains large gaps and several dense subregions.
To avoid any complication
due to methodological artefacts, we selected for frequency analysis
the simplest statistic available: the variance of
the least squares fit residuals divided by
the original variance of the data.
As the model to be  fitted we used a single
harmonic curve $\mu+A \cos(2 \pi t w_k)+
B \sin(2 \pi t w_k)$, where $\mu$ is the mean level of the light curve and
$w_k, (k=1,2,\dots,K)$ is one of the
trial frequencies. The computed row of $K$ variance ratios forms the
least squares (LS) frequency spectrum (see Pelt~\cite{pelt}).

The overall picture of the time variability of the star
in the range of periods $0.8-1000$ days is presented in Fig.~\ref{Fig1}.
For the channels V, B and U, most of the variability is well concentrated
around the zero frequency (the bunch of peaks around frequency $1$ cyc/day
is a result of aliasing from zero frequency). The relatively small
peak at $2\fd641$ in the U band is just the feature we are going to analyse
in more detail. As it is well seen from the same figure,
the strength of the peak for B--V
is significantly amplified, and for the U--V colour
(computed as (U--B)+(B--V)) the peak becomes the strongest
in the full range of frequencies. Its statistical significance can be
demonstrated using different methods. For instance, when we carried out
1000 Monte-Carlo runs in the range of periods 2--3 days, the reshuffled
data never showed peaks of this strength. For the full displayed range
this gives a false alarm probability which is less than $0.0075$. Due to
the bad spacing of the data, any more detailed analysis
can give only formally
better levels of significance. This is why we tried to
test the $2\fd641$ feature by using a careful analysis of separate subsets
of the full data. The peak at $1\fd609$ in U--V (see Fig.~\ref{Fig1})
is a result of aliasing (${1 \over 1.609} \approx 1- {1 \over 2.641}$),
and we do not consider it further.

\subsection{General and specific variability}

\begin{figure*}
\resizebox{\hsize}{!}{\includegraphics{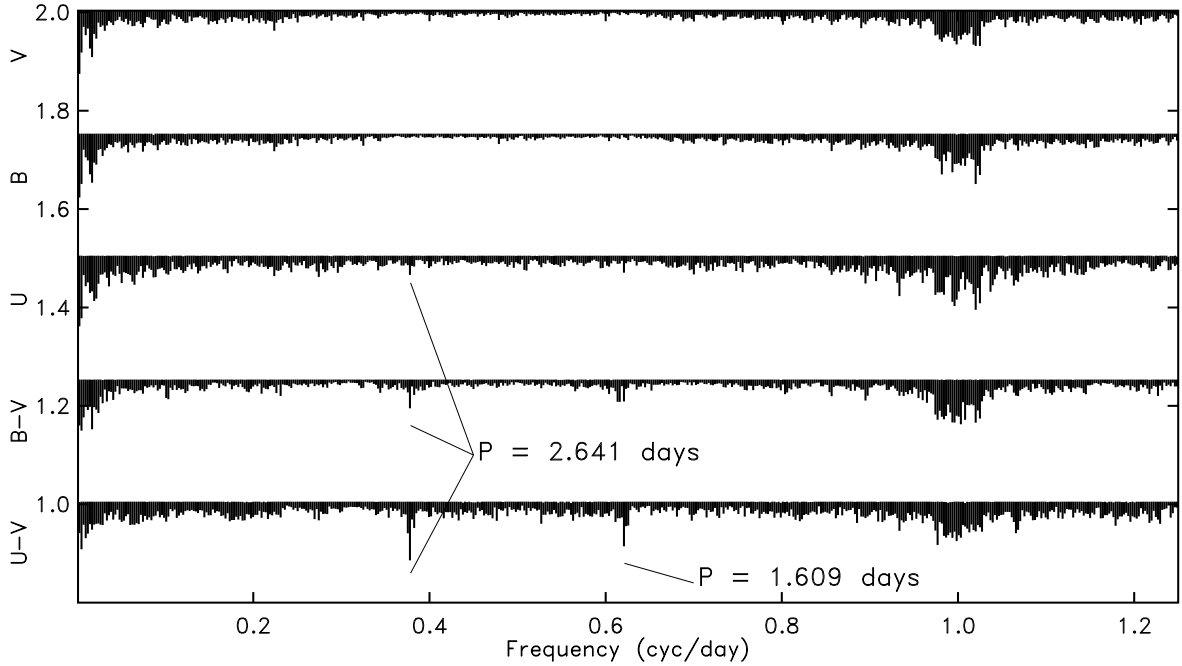}} 
\vspace*{-1cm}
\caption{Least squares spectra for magnitudes and colours.
Due to the long time base, the actual spectra contain
more than 300000 points each, so that the plot is somewhat schematic.
For every pixel of the plot, the minimum and maximum value of the spectra covered
by that pixel was computed, and the spectral content for the corresponding
pixel was plotted as a vertical line from minimum to maximum.}
\label{Fig1}
\end{figure*}

From the colour-magnitude diagrams (Fig.~\ref{col_mag}) we can conclude
that the total variability of RW~Aur~A consists of at least two components:
general variability (GV), which is a large-scale variability in all the
colours within about $0\fm5$, and
specific variability (SV), which can be best seen in the colour-magnitude
diagrams for U--B and B--V, as the scatter of the points around the linear trend.
This scatter is very small
in the red colours but steeply increasing towards the UV. Fig.~\ref{Fig1}
shows that the feature at $P=2\fd641$ belongs to SV (it is totally
absent from the V band). In order to
investigate the SV more closely, we should eliminate the linear trend in the
colour-magnitude diagrams and investigate the residual time variations in the
colours, which can be defined as
(B--V)$_{\rm res}$=(B--V)--$C\cdot$V+{\it const} for \hbox{B--V} and in a similar way for
other colours.
Thus, we included an extra term into the LS fit of the harmonics:
$\mu+A \cos(2 \pi t w_k)+B \sin(2 \pi t w_k)+C \cdot {\rm V}(t),$
and allowed coefficient $C$ to vary together with the other
model parameters $\mu,A,B$. As a result we obtained a new
statistic for the period search which allows to optimally separate
SV from GV.

\begin{figure}
\resizebox{\hsize}{!}{\includegraphics{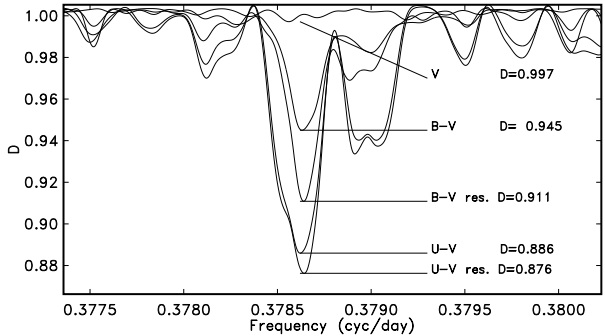}} 
\caption{Details of the LS spectra around period $P=2\fd641$.
Specific values of the test statistic (denoted here by D)
are also displayed. It is well seen that the
peaks are amplified in the spectra of the residual variances.}
\label{Fig2}
\end{figure}

\begin{figure}
\resizebox{\hsize}{!}{\includegraphics{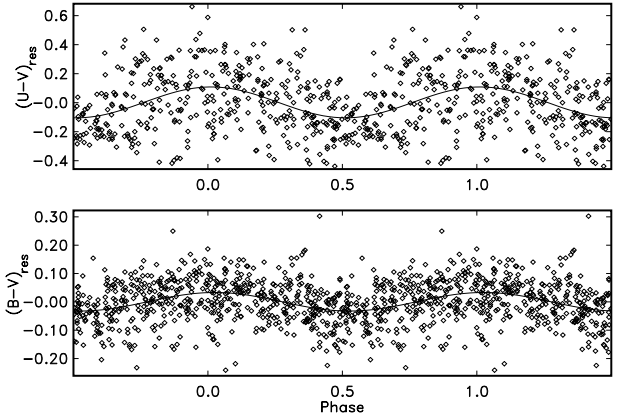}} 
\caption{The phase diagram for (U--V)$_{\rm res}$ and (B--V)$_{\rm res}$
with the period $2\fd641$.
The best fit harmonic is also given as a continuous line.}
\label{Fig3}
\end{figure}

The results of such a procedure can be
seen in Fig.~\ref{Fig2}, where short fragments of the LS
spectra are depicted. One can see that a strong improvement is obtained
for the B--V colour. The improvement is even more dramatic in the red
colours, where the amplitude of the periodic signal is very small
(SV$\ll$GV).
The phase diagrams with $P=2\fd641$
for the residual variations (U--V)$_{\rm res}$ and (B--V)$_{\rm res}$
are depicted in Fig.~\ref{Fig3}. The scatter
is still quite significant so that we need to look at the data
in more detail.

\begin{figure}
\resizebox{\hsize}{!}{\includegraphics{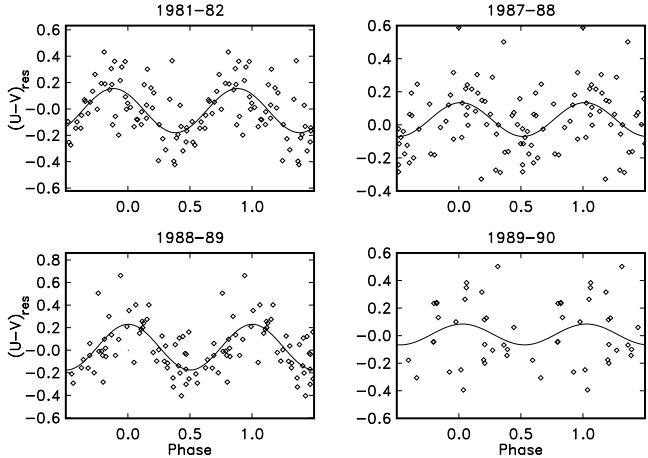}} 
\caption{Local fits for (U--V)$_{\rm res}$  with the period 2.641 days.
There are 59 points for the 1981--82 season, 54 points for 1987--88,
54 points for 1988--89, and 27 points for 1989--90.}
\label{Fig4}
\end{figure}

\subsection{Stability of the period}

Here, we will focus
on the analysis of the residual variation (U--V)$_{\rm res}$.
The largest number of observations
are concentrated in the following four seasons:
1981--82, 1987--88, 1988--89 and 1989--90. In Fig.~\ref{Fig4} we plotted
the results of the LS fits of a single harmonic with the period
$P=2\fd641$ for each of the four seasons separately.
The scatter for two seasons (1981--82 and 1988--89) is reasonably
small, but for season 1987--88 it is much larger, and for the last
season with the small number of points it is very high. However, the
amplitudes (maximum--minimum) are still significant ($0\fm33$, $0\fm20$,
$0\fm40$, $0\fm15$, respectively), especially if compared to the amplitude
$0\fm21$ of the full data fit (see Fig.~\ref{Fig3}). There is also a
reasonably good phase stability: the phase shifts around the full fit phase
are 20\%, 5.7\%, 3.3\%, 8.3\%, respectively.

As can be seen from Fig.~\ref{Fig2} the feature around the period $2\fd641$
consists of two main depressions in the LS frequency spectra. 
The smaller peak
right from the main peak ($P=2\fd639$) is a result of a 
cycle count ambiguity between
seasons 1981--82 and 1987--88. If we computed the LS
spectrum for the three subsets from 1981--1989 together, ignoring
all other data points, the peaks around
$2\fd641$ and $2\fd639$ occured to be comparably deep. Consequently, the last
digit in the value of the accepted period depends on
wildly scattered points outside the above mentioned seasons
and must be taken only as the best approximation.

In conclusion, we can say that the most prominent feature in the frequency
spectrum of the U--V colour is certainly a real regularity 
which is more or less persistent at least during eight years
(see Fig.~\ref{Fig4}).

\section{Discussion: Is there a hot spot?} \label{sec:discussion}

The periodic variations of the blue colours are most probably related to
the periodicities in the spectral lines discussed in Paper~I.  The spectroscopic
data set used in Paper~I covered only four years (1996--1999), which allowed 
to find a period within the range $2\fd6-2\fd9$.
The period of $2\fd77$ was selected as the best one for demonstration of the
phase diagrams in Paper~I, although with a period of $2\fd64$ the diagrams
look similar. In this paper, using the much larger time span of the photometric
data we confirm the period and prove that it was relatively stable over
at least several years.
The stability of the period probably excludes the possibility that the 
periodicity is caused by a local asymmetry in the magnetospheric structure, 
which is supposed to be a short-lived phenomenon. We still cannot distinguish 
between the binary hypothesis and the hypothesis of the inclined magnetic 
rotator (misalignment of the magnetic and rotational axes), discussed 
in Paper~I.

The amplitudes of the periodic colour variations rising steeply towards the
UV indicate the presense of a hot source of radiation (hot spot).
When the whole data set is used, the full amplitudes of the sinusoidal
oscillations are:
$0\fm21$ in U--V, $0\fm07$ in B--V, and about $0\fm02$ in V--R and V--I.
Using black body approximations for the fluxes from the photosphere and
the hot spot, one can estimate the temperature of the hot spot and the fraction
of the visible stellar disc covered by the spot (Vrba et al.~\cite{vrba93}).
With T$_{\rm phot}$=4800\,K, and the amplitudes given above,
we get T$_{\rm spot}$=15000\,K and the covering fraction 0.1\%.
Such a small spot gives only a $0\fm05$ amplitude in the V magnitude.

\begin{figure}
\centerline{\resizebox{7cm}{!}{\includegraphics{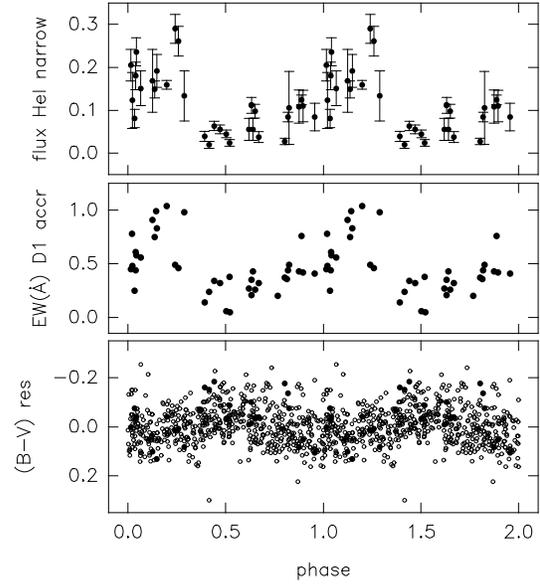}}}  
\caption{Phase diagrams with the period $2\fd641$.
Upper: flux (in~$10^{-12}$\,erg\,cm$^{-2}$\,s$^{-1}$) in the narrow HeI emission. 
The errors are due to uncertainties in the decomposition of the \ion{He}{i} emission profile 
into the narrow and broad components. 
Centre: equivalent width of the \ion{Na}{i} D$_1$ accretion component.
Lower: the residual B--V colour (filled circles: the photometry made simultaneously 
with the spectroscopy; open circles: the whole set of the photometric data).}
\label{new_3ph}
\end{figure}

In Paper~I we presented strong arguments in favour of the magnetospheric 
accretion model. The accelerated gas stream(s) flowing toward the star is 
clearly seen in the red-shifted components of many spectral lines. The 
strengths of the red-shifted components vary periodically, being increased
considerably at the phase when the accretion stream is in the line of sight. 
We found that the narrow \ion{He}{i} emission varies in correlation with
the red-shifted accretion components, and concluded that the narrow
emission originates from the shock region(s) at the bottom of the accretion 
stream, near the stellar surface. The broad emissions are believed to be formed 
in the global magnetosphere of the star. The two relevant phase diagrams are 
reproduced here with the new period $2\fd641$ and compared to the colour 
phase diagram (Fig.~\ref{new_3ph}).

\begin{figure}
\resizebox{\hsize}{!}{\includegraphics{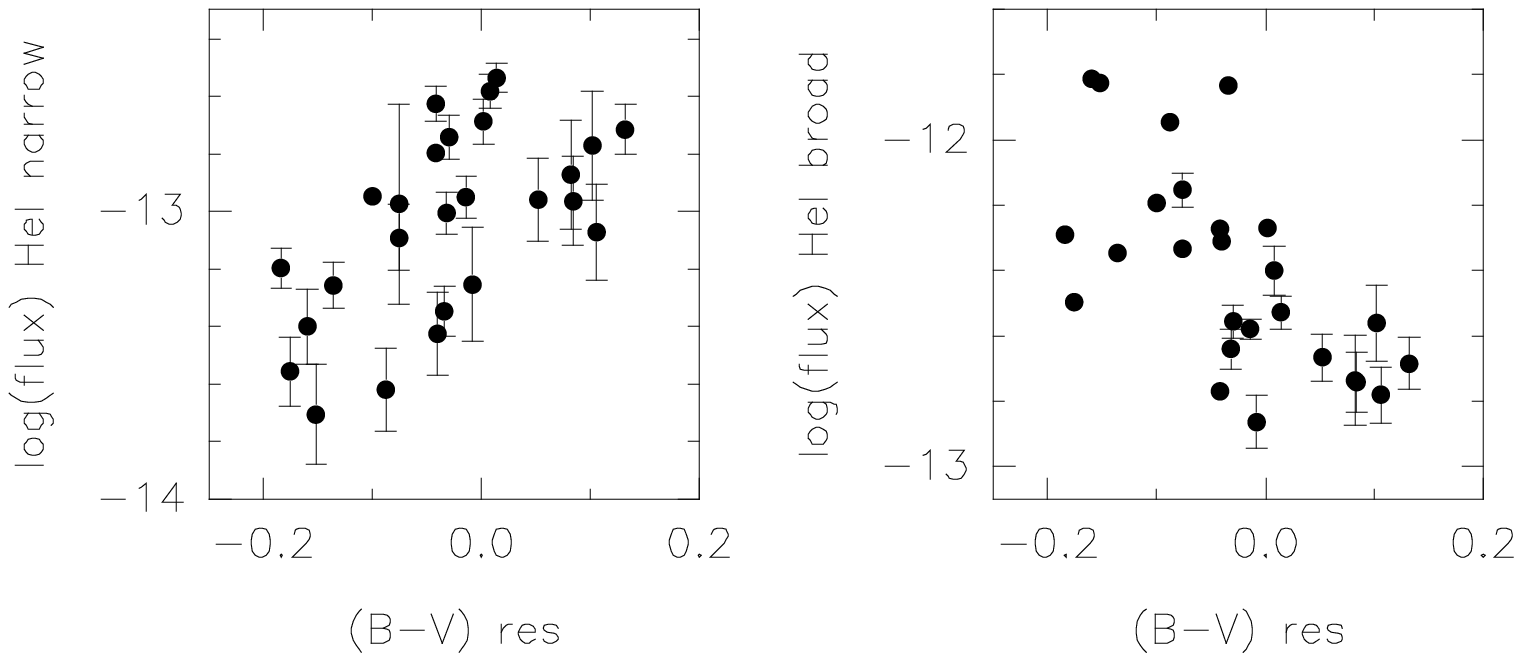}} 
\caption{Fluxes in the narrow and broad components of the \ion{He}{i}
emission vs. the residual B--V colours.}
\label{flux_bv}
\end{figure}

We might now identify the small hot spot with the accretion shock
at the stellar surface. Then we would expect the bluer colours when the narrow
\ion{He}{i} emission is stronger. However, from Fig.~\ref{new_3ph} one can
notice that quite the opposite is observed:
when the colours are bluer, the flux in the narrow \ion{He}{i} emission
is lower. This inverse correlation is better seen in the Fig.~\ref{flux_bv} with the
logarithmic scale for the flux. The "normal" correlation exists between the
colour and the {\it broad line} flux: the larger the broad line flux, the
bluer the colour (see also Paper I).

Hence, the colour varies in "anti-phase" to the strength of the accretion
components and to the flux in the narrow \ion{He}{i} emission,
in the sense that the star is redder when the shock region is facing the observer.
There is no doubt that magnetospheric accretion is going on in RW Aur A,
and that the accretion column(s) exists, but the expected hot spot at the base
of the accretion column does not reveal itself in our series of data. The small hot
spot, discussed above, may belong to the non-uniformely distributed hot gas,
which also radiates in the broad emission lines.


The apparent absence of the hot spot related to the accretion column
is really puzzling. Either the accretion shock is absent indeed,
or the complicated geometry of the flows and possible extinction of
light within the gas streams makes the observed variations so complicated.
Simultaneous spectroscopic, photometric and polarimetric monitoring
of the star could clarify the situation.

\begin{acknowledgements}
The authors thank Dr. Rudolf Duemmler for useful discussions and
careful reading of the manuscript, and the anonymous referee for
his/her valuable comments.
The work of J.P. was partly supported by the Estonian Science Foundation
(grant No. 4697).
\end{acknowledgements}

\end{document}